\documentclass[prd,twocolumn,tightenlines,superscriptaddress,nofootinbib,
showpacs]{revtex4}
\usepackage{amssymb,latexsym}
\usepackage{amsmath,amsbsy,bbm}
\usepackage{epsfig,bm}
\usepackage{graphicx,comment}
\usepackage{color}
\usepackage{soul}
\usepackage{csquotes}
\usepackage{url}
\usepackage[normalem]{ulem}
\begin{document}

\title{Quantum criticality at finite temperature for two-dimensional $JQ_3$ models on the square and the honeycomb lattices}

\author{J.-H. Peng}
\affiliation{Department of Physics, National Taiwan Normal University,
  88, Sec.4, Ting-Chou Rd., Taipei 116, Taiwan}
\author{D.-R. Tan}
\affiliation{Department of Physics, National Taiwan Normal University,
  88, Sec.4, Ting-Chou Rd., Taipei 116, Taiwan}
\author{L.-W. Huang}
\affiliation{Department of Physics, National Taiwan Normal University,
88, Sec.4, Ting-Chou Rd., Taipei 116, Taiwan}
\author{F.-J. Jiang}
\email[]{fjjiang@ntnu.edu.tw}
\affiliation{Department of Physics, National Taiwan Normal University,
88, Sec.4, Ting-Chou Rd., Taipei 116, Taiwan}

\begin{abstract}
  
  We study the quantum criticality at finite temperature for three two-dimensional (2D) $JQ_3$ models using
  the first principle nonperturbative quantum Monte Carlo calculations (QMC). In particular,
  the associated universal quantities are obtained and their inverse temperature dependence are
  investigated. The considered models are known to have quantum phase transitions from the N\'eel order to the valence bond solid.
  In addition, these transitions are shown to be of second order for two of the studied models, with the remaining one being of first order.
  Interestingly, we find that the outcomes obtained in our investigation are consistent
  with the mentioned scenarios regarding the nature of the phase transitions of the three investigated models.
  Moreover, when the temperature dependence of the studied universal quantities is considered, 
	a substantial difference between the two models possessing second order phase transitions and the remaining model is observed. 
	Remarkably, by using the associated data from both the models that may have continuous transitions, 
	good data collapses are obtained for a number of the considered 
	universal quantities. 
	The findings presented
  here not only provide numerical evidence to support the results established in the literature regarding the nature of the
  phase transitions of these $JQ_3$ models, but also can be employed as certain promising criterions
  to distinguish second order phase transitions from
  first order ones for the exotic criticalities of the $JQ$-type 
	models.
  Finally, based on a comparison between the results calculated here and the 
	corresponding theoretical predictions, we conclude that
  a more detailed analytic calculation is required in order to fully
  catch the numerical outcomes determined in our investigation.

\end{abstract}

\vskip-0.25cm

\maketitle

\section{Introduction}

Studying the characteristics of transitions between various phases of matters has long been an active
research topic in the condensed matter physics \cite{Nig92,Car96,Car10,Sac11}. In particular, with the advances of experimental techniques
and theoretical approaches, some exotic phases and (or) abnormal transitions between various states are
observed and (or) proposed. One such a noticeable example is the transition between the N\'eel phase and
the valence bond solid \cite{San07}.

The so-called $SU(2)$ $JQ$-type models of (quantum) spins in two-dimension (2D) \cite{San07,Lou09,Sen10},
which will be introduced later, are systems hosting the exotic phase transition between
the N\'eel and the valence bond solid states. The models have two types of
couplings, namely the $J$-term and the $Q$-term. In addition, when the magnitudes of the couplings increase, 
the $J$-term and the $Q$-term
favor the formation of singlets and pairs of singlets (valence bond solid, VBS),
respectively. On the one hand,
when the magnitude of $Q$ is much smaller than that of $J$, the $SU(2)$ symmetry
is broken spontaneously and the system is in the antiferromagnetic
long-range order. On the other hand, the translation symmetry is broken in the case of $Q \gg J$.
If one starts with a value of $Q/J$ ($Q >0$ and $J$ is fixed to be 1) so that the long-range order exists
and the translation symmetry of the system is unbroken, then by increasing the magnitude of $Q$
one expects that the restoration of the broken $SU(2)$ symmetry and the breaking
of the translation symmetry should occur at some couplings
$Q/J$. In particular, these two distinct physical incidents may take place at
different values of $Q/J$. If the two mentioned transitions occur at the same
$\left(Q/J\right)_c$, which is the most interesting scenario, then without fine-tuning this phase transition should be first order
according to the famous Landau-Ginsburg paradigm \cite{Gin50}.

Although the intriguing exotic phase transition introduced in the previous paragraph should be first order in general,
it is argue theoretically that this phase transition, as well as those which belong to some other models
possessing similar characteristics of the $JQ$-type models, can be second order \cite{Sen040,Sen041}.
Such an exotic critical behavior is referred to as the deconfined quantum criticality (DQC) in the literature. 

Many numerical investigations have been carried out in order to understand DQC, and various
conclusions are obtained \cite{Kuk06,Mel07,Jia08,Kuk08,San10,Che13,Har13}.
Although consistent critical exponents are obtained by simulating square lattices as large as
448$^2$ \cite{Sha16,San201}, one still cannot rule out the possibility that the considered phase transition
of the $SU(2)$ $JQ$ model (on the square lattice) is a weakly first order one. Indeed, it is shown
in Ref.~\cite{Iin19} that while the phase transition of the ferromagnetic 5-state Potts model on the square lattice
is a weakly first order phase transition, yet a high quality data collapse with certain values
of critical exponents, which should be a feature of a second order phase transition, is reached.
It is also unexpected that when more and more data of large lattices are included in the associated analyses, 
the determined values of $(Q/J)_c$ and the correlation length exponent $\nu$ change accordingly
in a sizeable manner \cite{San07,Mel07,San10,Sha16,San201}. For instance, when the largest
linear system sizes are $L=32$, 64 (128), 256, and 448, the calculated values of $\nu$ are given by
0.78, 0.68, 0.446, and 0.448 (or 0.455), respectively.
Hence, it will extremely interesting to
explore further the issue of the nature of these exotic phase transitions.  
In conclusion, at the moment it is a consensus that if these exotic phase transitions are second order,
then they must receive abnormally large corrections (to the scaling at the corresponding criticalities).

Studies of the $SU(2)$ $JQ$-type models have been targeted at uncovering the nature of DQC. Moreover, due to
the famous Mermin-Wagner theorem \cite{Mer66,Hoh67,Col73,Gel01},
investigating the finite-temperature properties of these models are focusing on the VBS perspective.
Yet by exploring the quantum critical regime (QCR) and the associated universal quantities, which are
features of the antiferromagnetic phase of the $JQ$-type
models \cite{Chu93,Chu931,Chu94,San95,Tro96,Tro97,Tro98,Kim99,Kim00,San11,Sen15,Tan181},
one may gain certain aspects of the nature of DQC \cite{Kau08,Kau081,Puj13}. Indeed, studying the QCR of various $JQ$-type models which have either
a first order or a second order phase transitions may shed some light on the
DQC.

Due to the motivations described in the previous paragraphs, in this investigation we study three 2D $SU(2)$ $JQ_3$ models (Which will be introduced in detail 
later). In particular, two of these three systems are on the square lattice and the remaining one is on the honeycomb
lattice \cite{Lou09,Sen10,Puj13,Har13}. Moreover, among these three considered $JQ_3$ models,
one is shown to have a first order phase transition and the other two are argued to possess second order phase
transitions \cite{Lou09,Sen10,Puj13,Har13}. As a result, studying these three
systems may reveal discrepancies of the features of QCR that result from the nature of the phase transitions.

In our calculations, the associated universal quantities of QCR, namely $S(\pi,\pi)/\left(\chi_s T\right)$, $\chi_u c^2/T$ and $\rho_s L /c$
are obtained. Here $S(\pi,\pi)$, $\chi_s$, $T$, $\chi_u$, $c$, $\rho_s$, and $L$ are the staggered structure factor,
the staggered susceptibility, the temperature, the uniform susceptibility, the spinwave velocity, the spin stiffness, and the linear system size,
respectively.
Remarkably, by simulating large systems
($448^2$ spins for the square lattice and more than $300^2$ spins for the honeycomb lattice), we find that there is a substantial difference in the temperature dependence of these considered universal quantities between models with first and second order phase transitions. Specifically,
the features of QCR only show up for the two $JQ_3$ models which are shown to possess second order phase transitions in the literature.
In addition, the $S(\pi,\pi)/\left(\chi_s T\right)$ data of both the models having second order phase transitions reach the same plateau when regarded as a function of the inverse temperature $\beta$. A similar scenario occurs for $\chi_u c^2/T$ as well.

Remarkably, by considering the $S(\pi,\pi)/\left(\chi_s T\right)$ data of both the models having the characteristics of QCR as a function of $\beta c$, a high quality data collapse shows up. The quantity $\chi_u c^2/T$ also demonstrates the same behavior when treated as a function of $\beta c$. 
The outcome associated with $\rho_sL/c$ for the model on the honeycomb lattice requires a careful interpretation and
can be satisfactory accounted for by taking into account a logarithmic correction to the corresponding theoretical predictions.

Although one cannot completely rule out the possibility of weakly first order phase transitions, the obtained results shown here provide evidence to support the hypothesis that the targeted phase transitions, from the N\'eel to the valence bond solid states, are likely continuous for two of the three studied $JQ_3$ models.

Finally, we also make a comparison between the numerical results obtained here and the related theoretical predictions
in Refs.~\cite{Chu94} and \cite{Kau08}, and have concluded that the leading 
order large $N$-expansion calculations carried out in Refs.~\cite{Chu94}
and \cite{Kau08} are not sufficient to match the outcomes determined in our 
study.

The findings presented in our investigation are not only
interesting in themselves, but also can be adopted as useful criterions to distinguish first order phase transitions from the second order ones
associated with DQC.

The rest of this paper is organized as follows. In Sect. II
the studied $JQ_3$ models and the relevant observables of QCR
are introduced. We then demonstrate our numerical results, particularly
the mentioned dramatical difference of the temperature dependence of
the calculated universal quantities are demonstrated in Sec. III.
Finally, in Sec. IV we present discussions and conclusions.

\begin{figure}
\vskip-0.5cm
\begin{center}
    \includegraphics[width=0.5\textwidth]{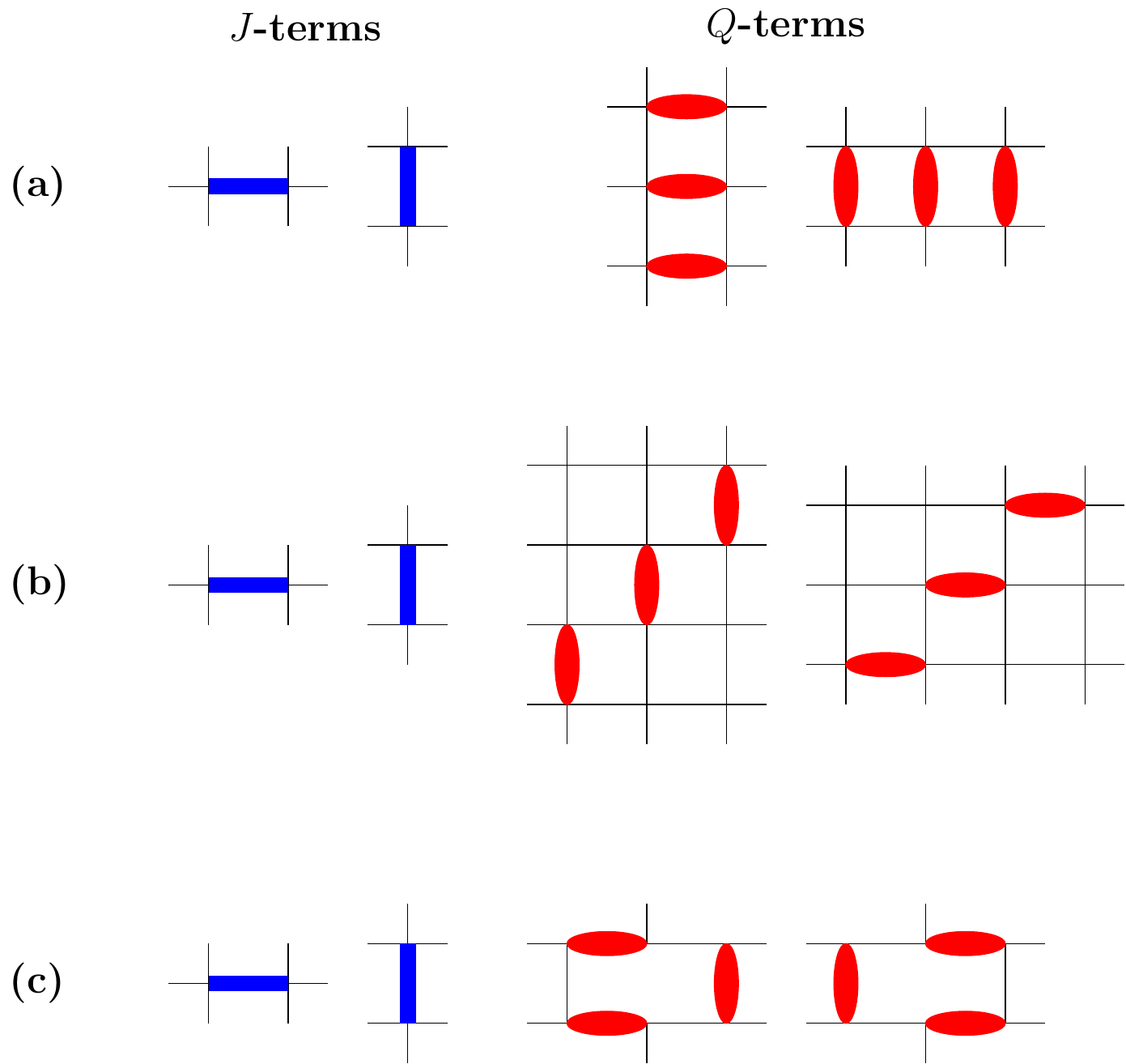}\vskip0.25cm
\end{center}\vskip-0.5cm
\caption{Three two-dimensional (2D) $JQ_3$ models considered in this study.
For model (b), two more $Q$-terms, which can be obtained by a 90 degree rotation of those depicted above, are not shown here.}
\label{fig1}
\end{figure}

\section{Microscopic models and observables}

The Hamiltonians of the three investigated 2D $SU(2)$ $JQ_3$ models on the square and the honeycomb lattices
have the same expression which is given by
\begin{eqnarray}
\label{hamilton}
H &=& J\sum_{\langle ij \rangle}\vec{S}_i\cdot\vec{S}_j - Q\sum_{\langle ijklmn\rangle}P_{ij}P_{kl}P_{mn},\nonumber \\ 
P_{ij} &=&\frac{1}{4}-\vec{S}_i\cdot\vec{S}_j,
\end{eqnarray}
where in Eq.~(1) $J$ (which is set to be 1 here) and $Q$ are the
couplings for the two-spin and the six-spin interactions, respectively,
$\vec S_{i} $ is the spin-1/2 operator at site $i$, and
$P_{ij}$ is the singlet pair projection
operator between nearest neighbor sites $i$ and $j$.
Figure~\ref{fig1} contains the cartoon representations of the studied models.
In the following, the top, the middle, and the bottom models of
fig.~\ref{fig1} will be named the ladder, the staggered, and the honeycomb $JQ_3$ models
respectively, if no confusion arises. We would also like to emphasize the fact that
although the boundary condition (BC) for the honeycomb lattice considered here is different from that
of Ref.~\cite{Puj13}, bulk properties such as the critical point should be independent of
the implemented BC. Hence, whenever $(Q/J)_c$ for the $JQ_3$ model on the honeycomb lattice
is needed, the one determined in Ref.~\cite{Puj13} will be used.

To study the quantum critical regime associated with the investigated $JQ_3$ models,
particularly to calculate the corresponding universal quantities, several relevant
observables, such as the staggered structure factor $S(\pi,\pi)$,
the staggered susceptibility $\chi_s$,
the uniform susceptibility $\chi_u$, the spinwave velocity $c$, the spin stiffness $\rho_s$, and the temporal and
spatial winding numbers squared ($\langle W_t^2\rangle$ and $\langle W_i^2 \rangle$ for $i=1,2$) are measured in our calculations.
The formal expressions of these observables (on a $L_1$ by $L_2$ lattice) are as follows

\begin{eqnarray}
  S(\pi,\pi) &=& \frac{3}{L_1L_2}\langle \left(m_s^{z}\right)^{2} \rangle,\\
  m_s^{z} &=& \sum_i (-1)^{i_1+i_2}S_i^z,\\
  \chi_s &=& \frac{3}{L_1L_2}\int_0^{\beta} \langle m_s^z(\tau) m_s^z(0)\rangle d\tau,\\
  \chi_u &=& \frac{\beta}{L_1L_2}\langle \left(m^z\right)^2\rangle, \\
  m^z&=& \sum_{i}S_i^z,\\
	\rho_s &=& \frac{3}{4\beta}\left(\langle W_1^2\rangle + \langle W_2^2 \rangle\right),
\end{eqnarray}
where $\beta$ and $L_i$ for $i=1,2$ are the inverse temperature and the linear
box size (in the $i$-spatial direction) used in the simulations, respectively,
and $S_i^z$ is the third-component of
the spin-half operator $\vec{S}_i$ at site $i$.

\section{The numerical results}

To calculate the desired physical observables, 
we have carried out large-scale quantum Monte Carlo calculations (QMC) 
using the stochastic series expansion (SSE) algorithm with very efficient 
operator-loop update \cite{San99,San101}. In addition, the simulations
are performed at (close to) the associated critical points established in the
literature. Specifically, we conduct the simulations at
$(Q/J)_c = 1.500$, 1.1933 and 1.1936 for the ladder, the staggered, and the honeycomb
$JQ_3$ models, respectively. Finally, outcomes of
up to $448^2$ (over $300^2$) spins on the square (honeycomb) lattice are
obtained.

The calculated numerical results
regarding the universal quantities introduced previously will be described in detail in the following.
Particularly, in the analyses we assume that the associated dynamic critical exponents $z$ are all given by
$z=1$ \cite{Sen040,Sen041}.

\subsection{$S(\pi,\pi)/\left(\chi_sT\right)$}

\begin{figure}
\vskip-0.5cm
\begin{center}
  \vbox{
    \includegraphics[width=0.45\textwidth]{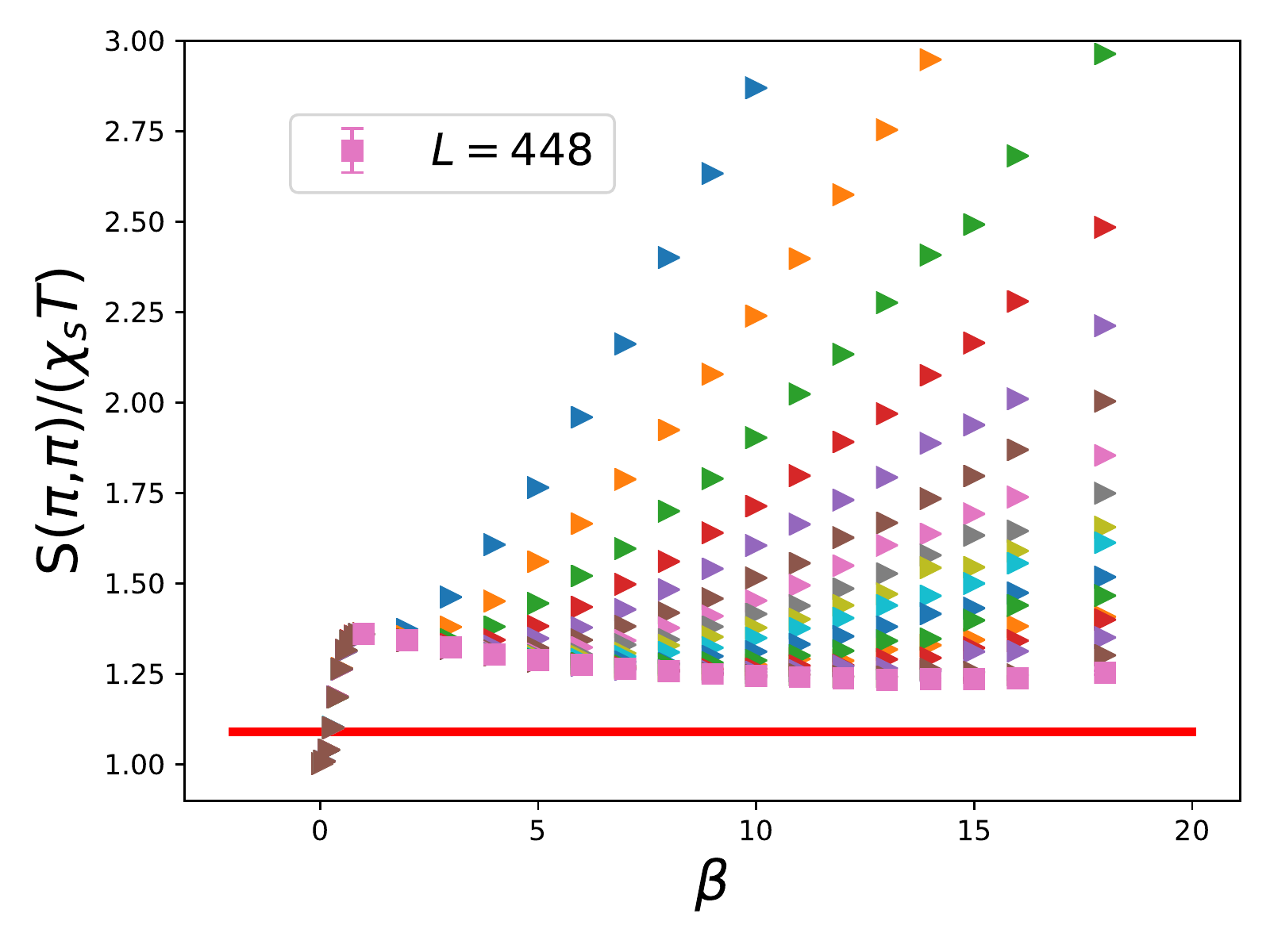}\vskip0.25cm
\includegraphics[width=0.45\textwidth]{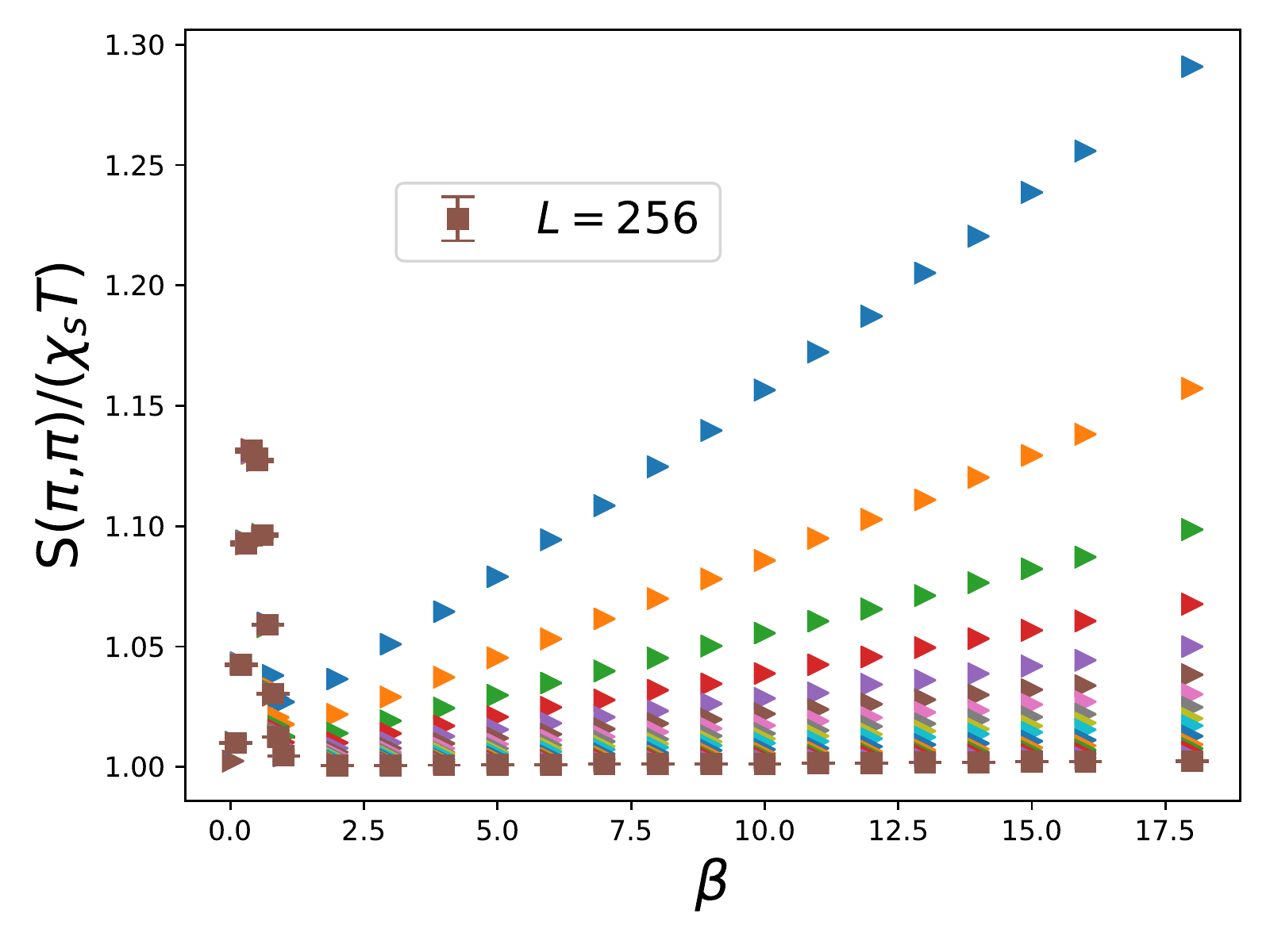}
}
\end{center}\vskip-0.5cm
\caption{$S(\pi,\pi)/\left(\chi_s T\right)$ as functions
  of $\beta$ for various $L$ for the ladder (top) and the staggered (bottom) $JQ_3$ models.}
\label{spipi1}
\end{figure}

\begin{figure}
\vskip-0.5cm
\begin{center}
    \includegraphics[width=0.45\textwidth]{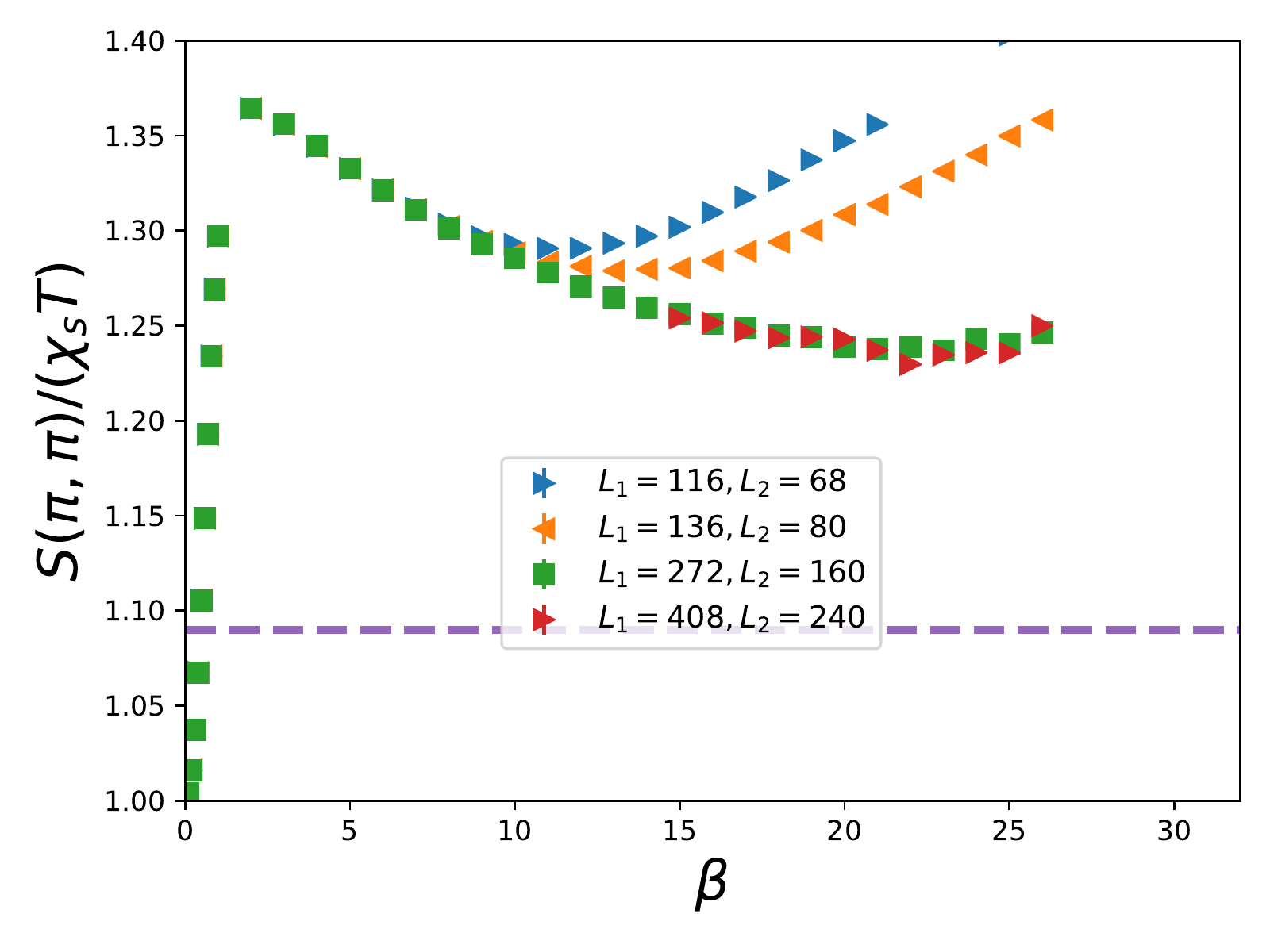}\vskip0.25cm
\end{center}\vskip-0.5cm
\caption{$S(\pi,\pi)/\left(\chi_s T\right)$ as functions
  of $\beta$ for various $L$ for the honeycomb $JQ_3$ models.}
\label{spipi2}
\end{figure}

The observable
$S(\pi,\pi)/\left(\chi_s T\right)$ as functions of $\beta$ for $L$ = 24, 32, 40,..., 192, 256, and 448 (from top to bottom) for the ladder and the staggered (only up to $L$=256) $JQ_3$ models are demonstrated in 
fig.~\ref{spipi1}. 
Similarly, fig.~\ref{spipi2} contains the results of $S(\pi,\pi)/\left(\chi_sT\right)$ versus $\beta$ for the honeycomb $JQ_3$ model.

As can been seen from the top panel of fig.~\ref{spipi1} which is associated
with the ladder $JQ_3$ model, although mild
(finite) temperature and system size dependence may still be there for the
largest lattice,
it is clear that the
$S(\pi,\pi)/\left(\chi_sT\right)$ of $L=448$ reaches a (more or less) plateau,
i.e., a constant value close to 1.25, for $\beta > 4 $. This is without doubt a character of the
quantum critical regime. The horizontal solid line shown in the figure is 1.09 and is the analytic result of this quantity (Ref.~\cite{Chu94}).
It is obvious that a more detailed
theoretical calculation is essential to catch the numerical results obtained here.

The $\beta$ dependence of $S(\pi,\pi)/\left(\chi_sT\right)$
for the staggered $JQ_3$ model is demonstrated in the bottom panel of fig.~\ref{spipi1}.
The results in the figure indicate $S(\pi,\pi)/\left(\chi_sT\right)$ reaches
the value of 1 rapidly, both at high and moderate low temperatures. Moreover, no sign
of the appearance of a plateau implies the absence of QCR. While further verification is
on request, most likely this is a feature of a first order phase transition.

Similar to the outcomes of the ladder $JQ_3$ model (on the square lattice),
the quantity $S(\pi,\pi)/\left(\chi_sT\right)$ of the honeycomb $JQ_3$ model
reaches a plateau for $\beta > 15$, see fig.~\ref{spipi2}.

It is interesting to notice that 
the results shown in figs.~\ref{spipi1} and \ref{spipi2} imply that
when compared with the ladder model,
larger values of $\beta$ are required for the quantity $S(\pi,\pi)/(\chi_s T)$
of the honeycomb $JQ_3$ model to saturate to its plateau.

\begin{figure}
\vskip-0.5cm
\begin{center}
    \includegraphics[width=0.45\textwidth]{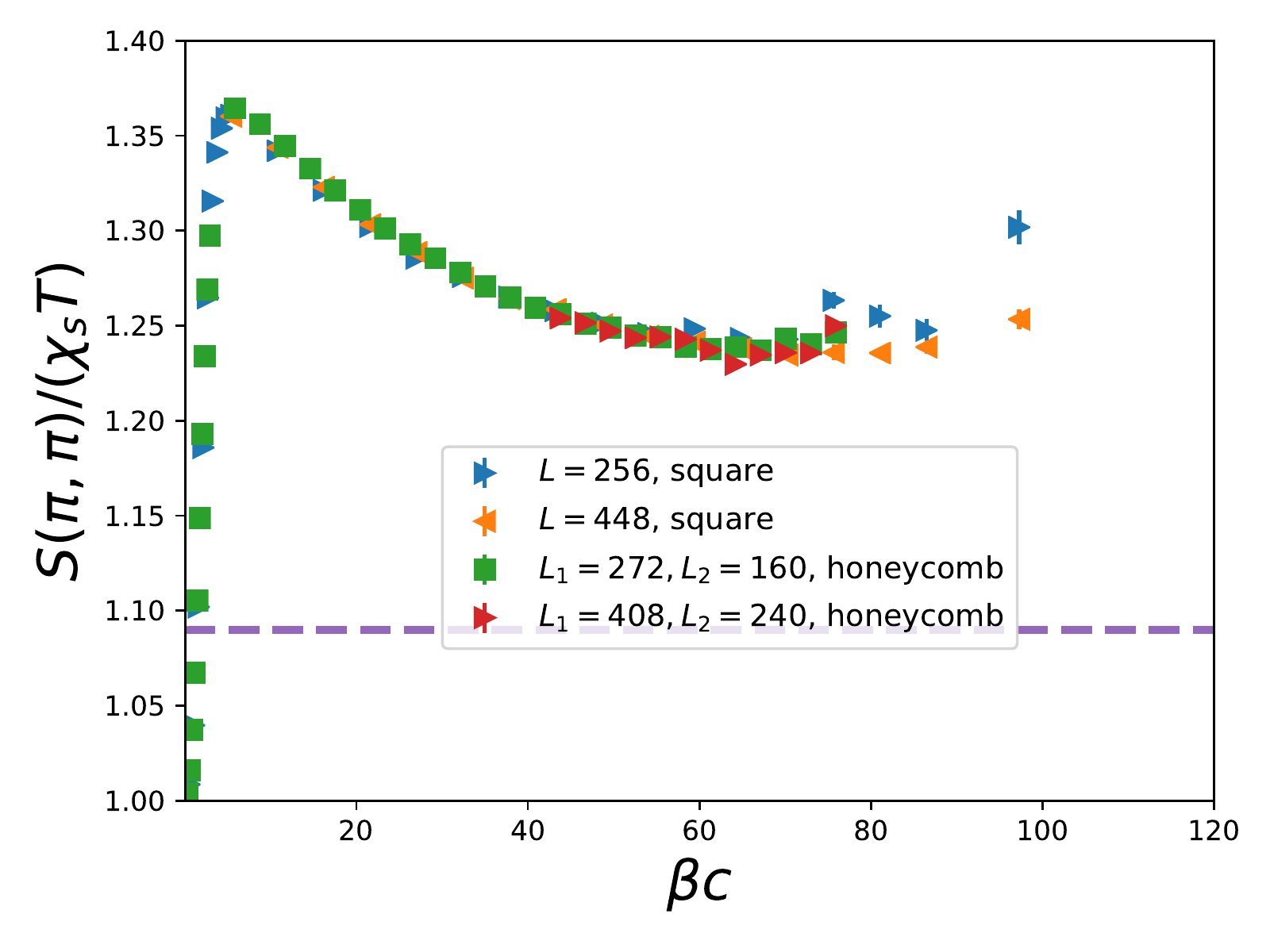}\vskip0.25cm
\end{center}\vskip-0.5cm
\caption{$S(\pi,\pi)/\left(\chi_s T\right)$ as functions
  of $\beta c$ for the ladder and the honeycomb $JQ_3$ models.}
\label{spipi3}
\end{figure}

Remarkably, the plateau for both the ladder and the honeycomb $JQ_3$
models take the same value. In particular, if the $S(\pi,\pi)/(\chi_s T)$
data on large lattices of both models are plotted as functions of
$\beta c$ (The calculations of $c$ will be detailed shortly),
a good data collapse outcome is obtained, see fig.~\ref{spipi3}.
This indicates it is highly probable that
QCR is indeed found in these two models and such an observation
in turn provides an evidence for
the associated phase transitions to be continuous.

The scenarios reached here are consistent with
the known outcomes in the literature regarding the exotic phase transitions
of these three investigated $JQ_3$ models.

\subsection{$\chi_u c^2/T$}

\subsubsection{Determination of the spinwave velocity $c$}

Since the low-energy constant spinwave velocity $c$ is required
for calculating the numerical value of $\chi_u c^2/T$, we have 
determined $c$ firstly.

The method used here for the calculations of $c$ is the
winding numbers squared \cite{Kau08,Sen15,Jia111,Jia112}. Specifically, for each considered
box size $L$, the $\beta$ is adjusted in the simulations so that three winding numbers
squared $\langle W_i^2\rangle$ for $i=1,2$ and $\langle W_t^2\rangle$ take
the same value. When this condition is met, the $c$ corresponding to the
simulated $L$ is $L/\beta$.

It should be pointed out that on the honeycomb lattice, square-shape
area can only be reached with certain aspect ratios $L_1/L_2$. As a result,
the simulations associated with the honeycomb lattice are performed
on these aspect ratios (of $L_1/L_2)$ so the (almost) square-shape area is obtained.

Finally, the calculations are done at $Q/J$ = 1.4925, 1.1923, 1.19 for the ladder, the staggered,
and the honeycomb $JQ_3$ models, respectively. 

\begin{figure}
\vskip-0.5cm
\begin{center}
  \vbox{
    \includegraphics[width=0.45\textwidth]{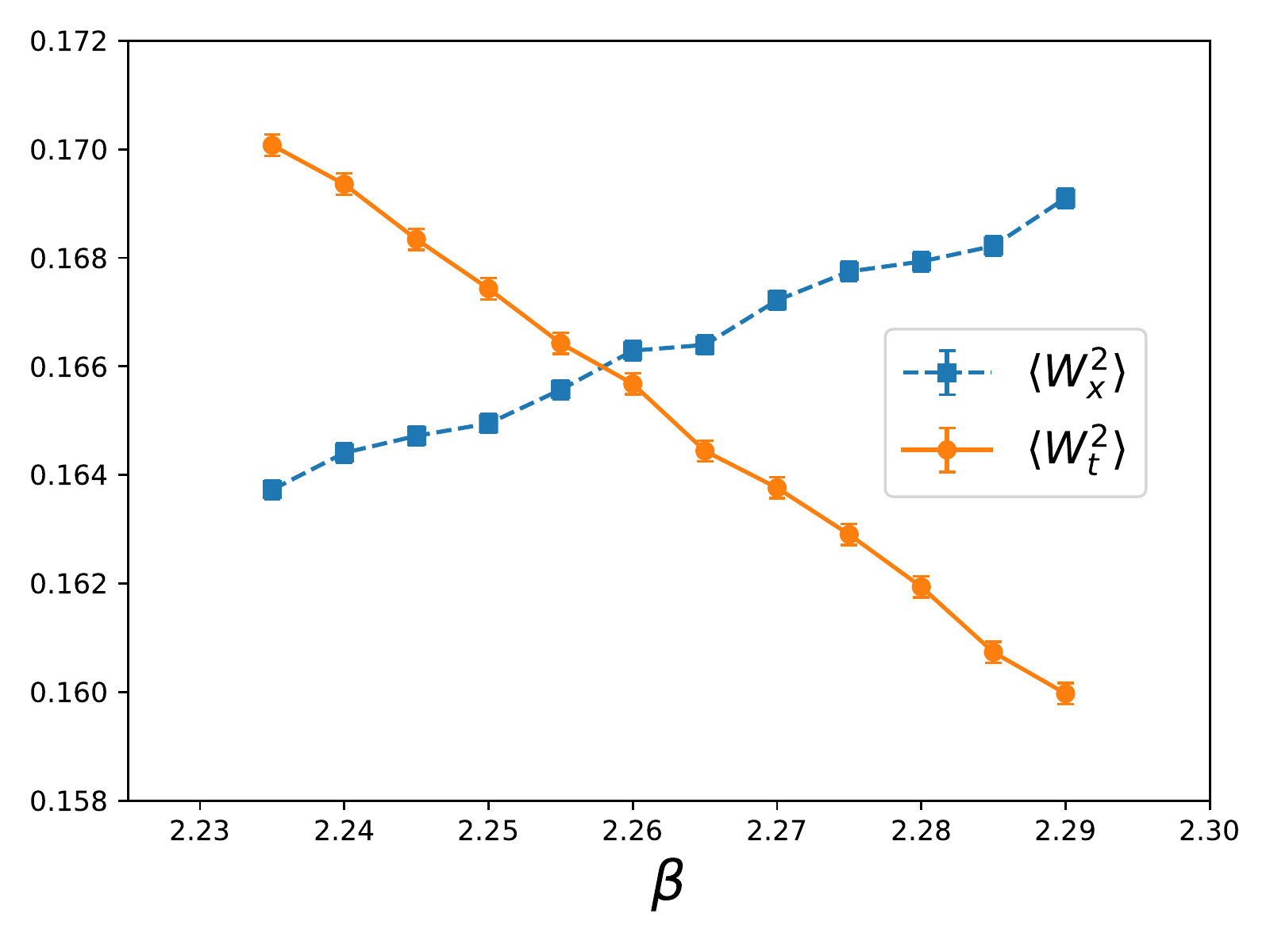}\vskip0.25cm
\includegraphics[width=0.45\textwidth]{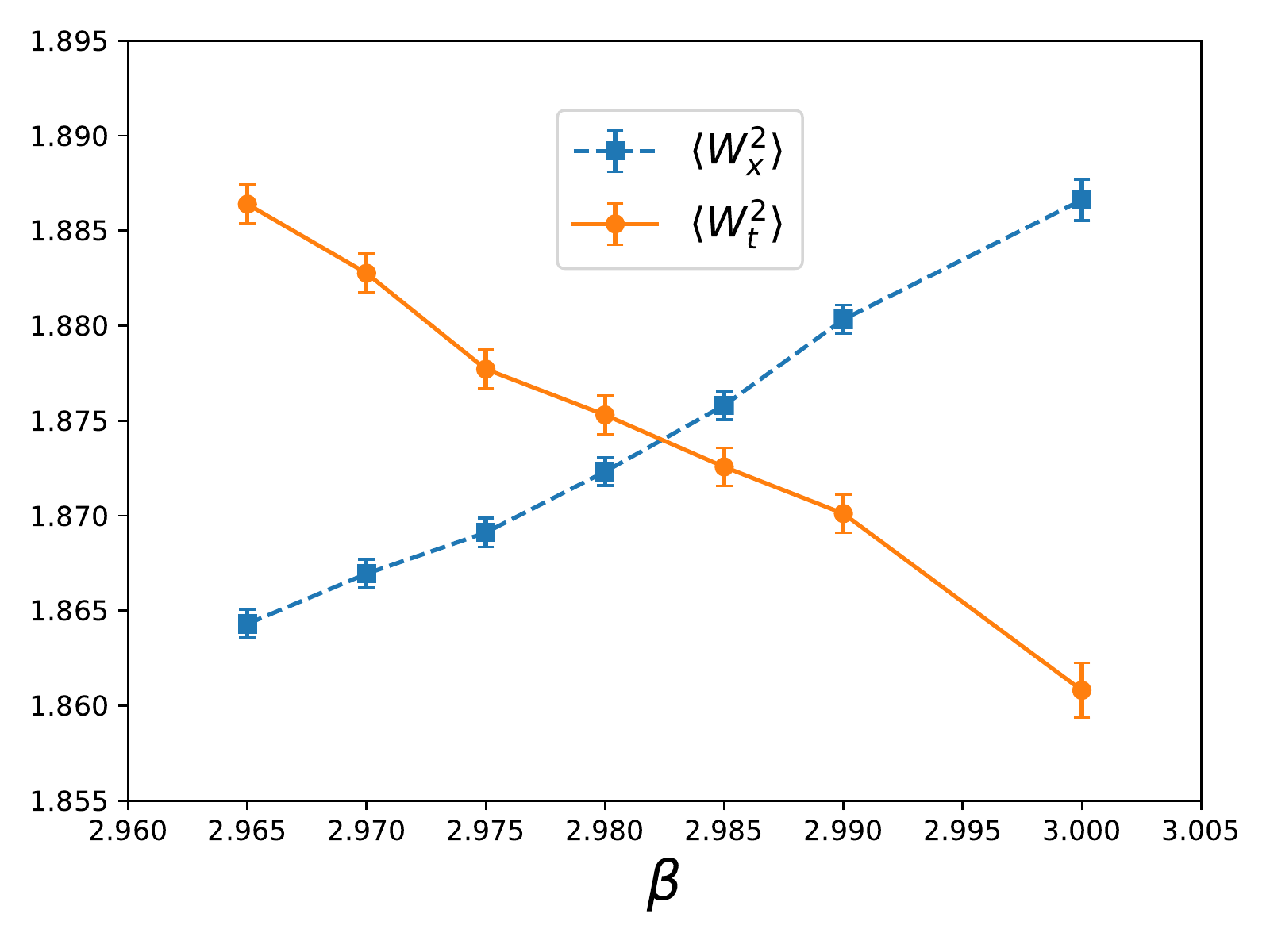}
}
\end{center}\vskip-0.5cm
\caption{$\langle W_x^2 \rangle$ and $\langle W_t^2 \rangle$ as functions
  of $\beta$ for the ladder (top, $L$=12) and the staggered (bottom, $L$=20) $JQ_3$ models.
  Here $\langle W_x^2 \rangle$ = $0.5\left(\langle W_1^2 \rangle+\langle W_2^2 \rangle\right)$.}
\label{velocity1}
\end{figure}

\begin{figure}
\vskip-0.5cm
\begin{center}
    \includegraphics[width=0.45\textwidth]{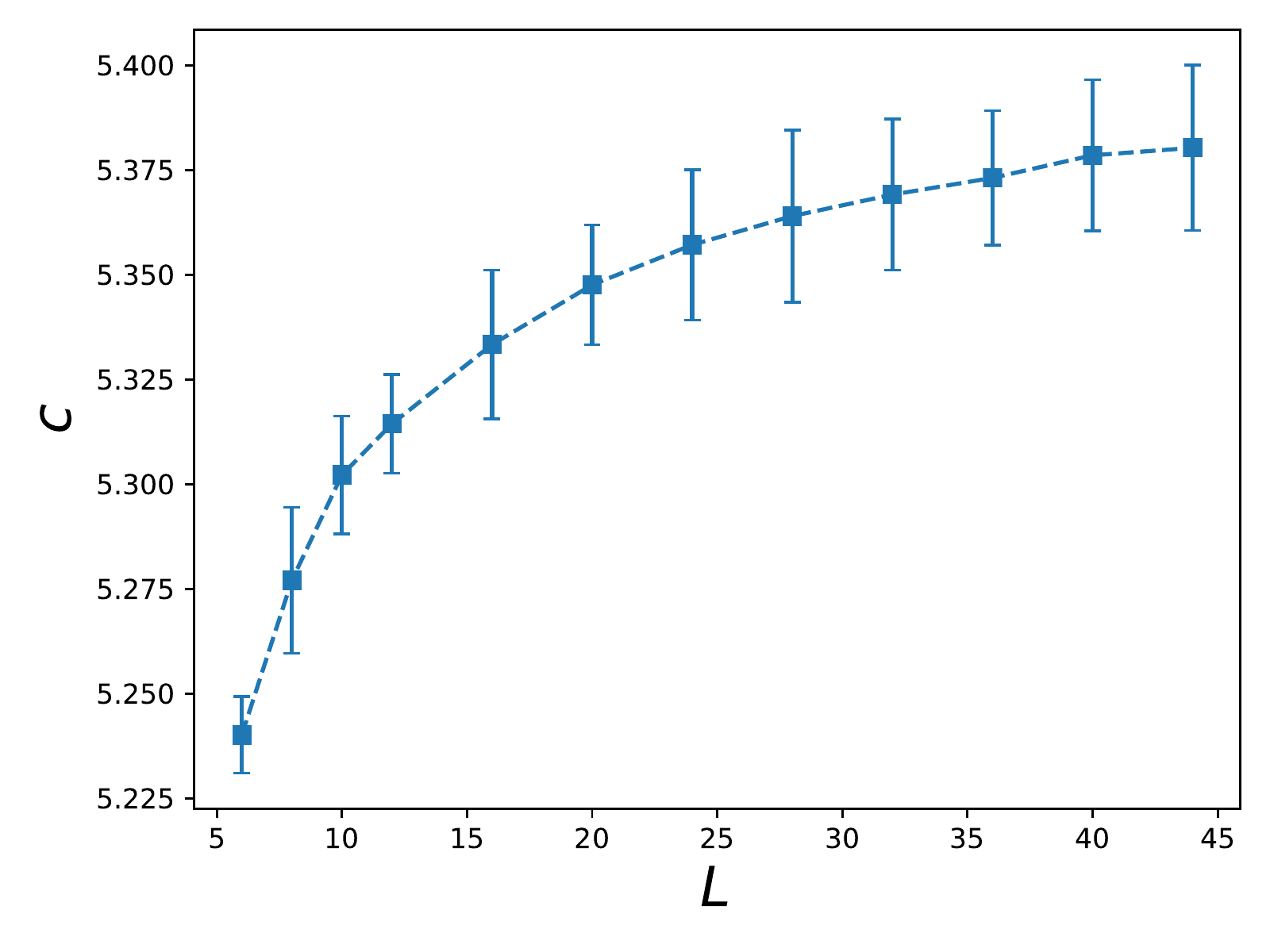}\vskip0.25cm
\end{center}\vskip-0.5cm
\caption{$c$ as a function
  of $L$ for the ladder $JQ_3$ models.}
\label{velocity2}
\end{figure}

The spatial and the temporal winding numbers squared as functions of
$\beta$ for the ladder and the staggered $JQ_3$ models are shown in the
top ($L=12$) and the bottom ($L=20$) panels of fig.~\ref{velocity1}.
In addition, the estimated $c$ related to various $L$ for the ladder $JQ_3$
model is demonstrated in fig.~\ref{velocity2}. A fit of applying the ansatz $c+a/L+b/L^2$ ($c+a/L+b/L^2+d/L^3$)
to the data of fig.~\ref{velocity2} leads to $c = 5.404(12)$ ($c=5.397(26)$).

Here we do not conduct the associated fits for the staggered $JQ_3$ model.
This is because as we will show shortly, with the fact that $c$ is a constant, the related $\chi_u c^2/T$ does not possess the expected
feature of QCR. As a result, calculating the bulk $c$ for this model is no long needed.   

Regarding the $c$ of the honeycomb $JQ_3$ model, the outcome 
on a lattice of $L_1 = 136$ and $L_2=80$ is quoted here and is given by
$c\sim 2.92$. We have also carried out simulations on a $34 \times 20$ honeycomb
lattice and the resulting $c$ is estimated to be $c = 2.874(24)$.
Since $2.874(24)/2.92$ is over 0.97 and the ratio of the linear box sizes of
these two honeycomb lattices is $\sqrt{136 \times 80}/\sqrt{34\times 20}$ = 4, 
one expects negligible finite-lattice effects for the quoted value of
$c \sim 2.92$.

\subsubsection{Determination of $\chi_u c^2/T$}

After determining the values of the spinwave velocity $c$ for all the three investigated $JQ_3$ model,
we turn to studying the universal quantity $\chi_u c^2/T$.

\begin{figure}
\begin{center}
    \includegraphics[width=0.45\textwidth]{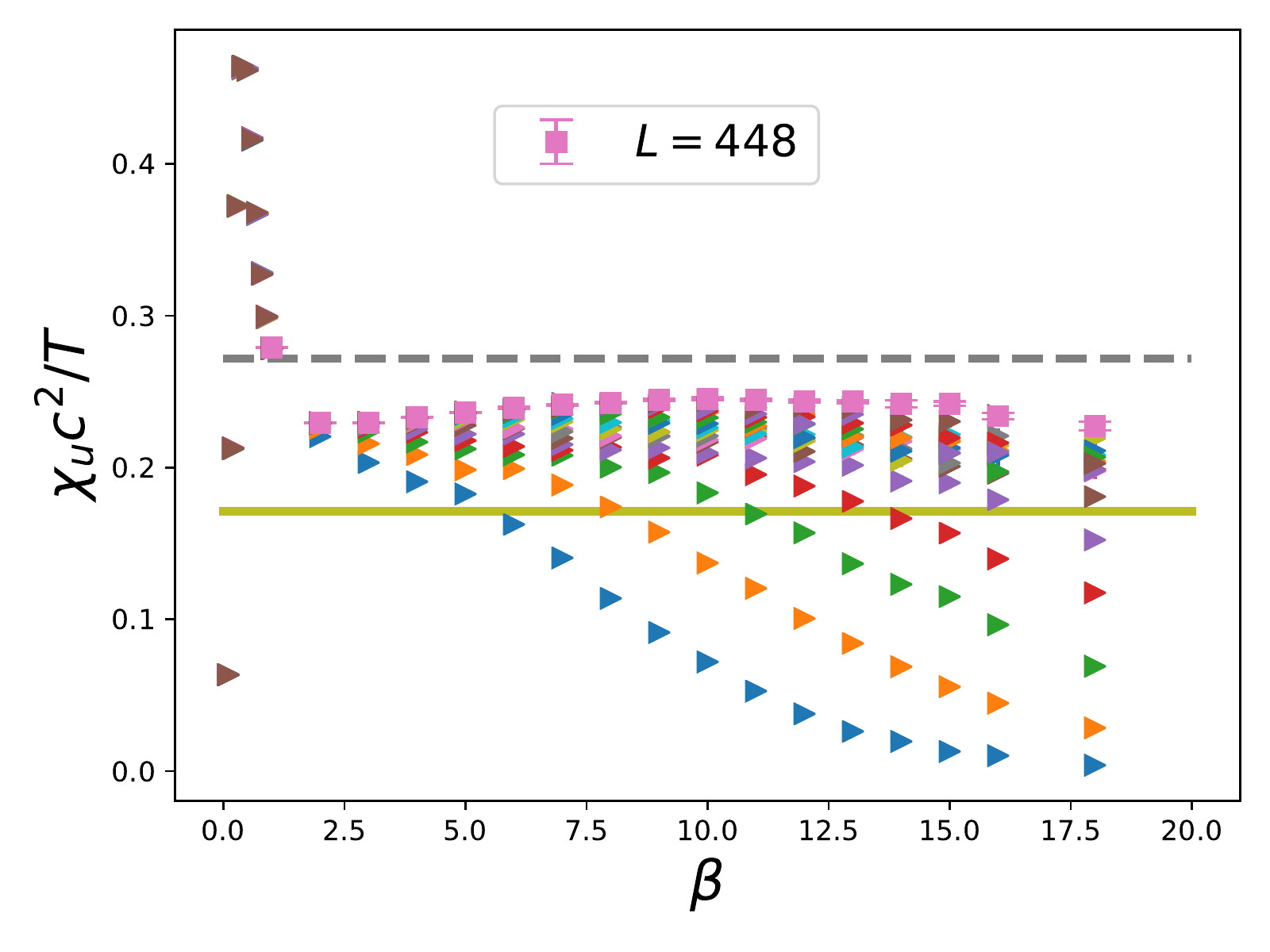}\vskip0.25cm
\end{center}\vskip-0.5cm
\caption{$\chi_u c^2 /T$ as functions
  of $\beta$ for various $L$ for the ladder $JQ_3$ models. The horizontal solid and dashed lines represent
the outcomes of 2.7185 and 1.7125 determined in Refs.~\cite{Chu94} and \cite{Kau08}, respectively.}
\label{chiu1}
\end{figure}

$\chi_u c^2/T$ as a function of $\beta$ for the ladder $JQ_3$ model is shown in fig.~\ref{chiu1}.
The figure suggests convincingly that the observable $\chi_u c^2/T$ of this model reaches a plateau for $\beta \ge 2$,
which is a feature of QCR. Similar to that of the quantity $S(\pi,\pi)/\left(\chi_s T\right)$,
this characteristic provides yet another evidence that the corresponding phase transition
is likely a second order one. The horizontal solid (2.7185) and dashed (1.7125) lines in fig.~\ref{chiu1} represent
the theoretical calculations obtained in Refs.~\cite{Chu94} and \cite{Kau08}, respectively. It is clear that
the next to leading order large $N$-expansion computations
carried out in these references are not sufficient to match the numerical value of $\chi_u c^2/T$ determined in this study.

\begin{figure}
\vskip-0.5cm
\begin{center}
    \includegraphics[width=0.45\textwidth]{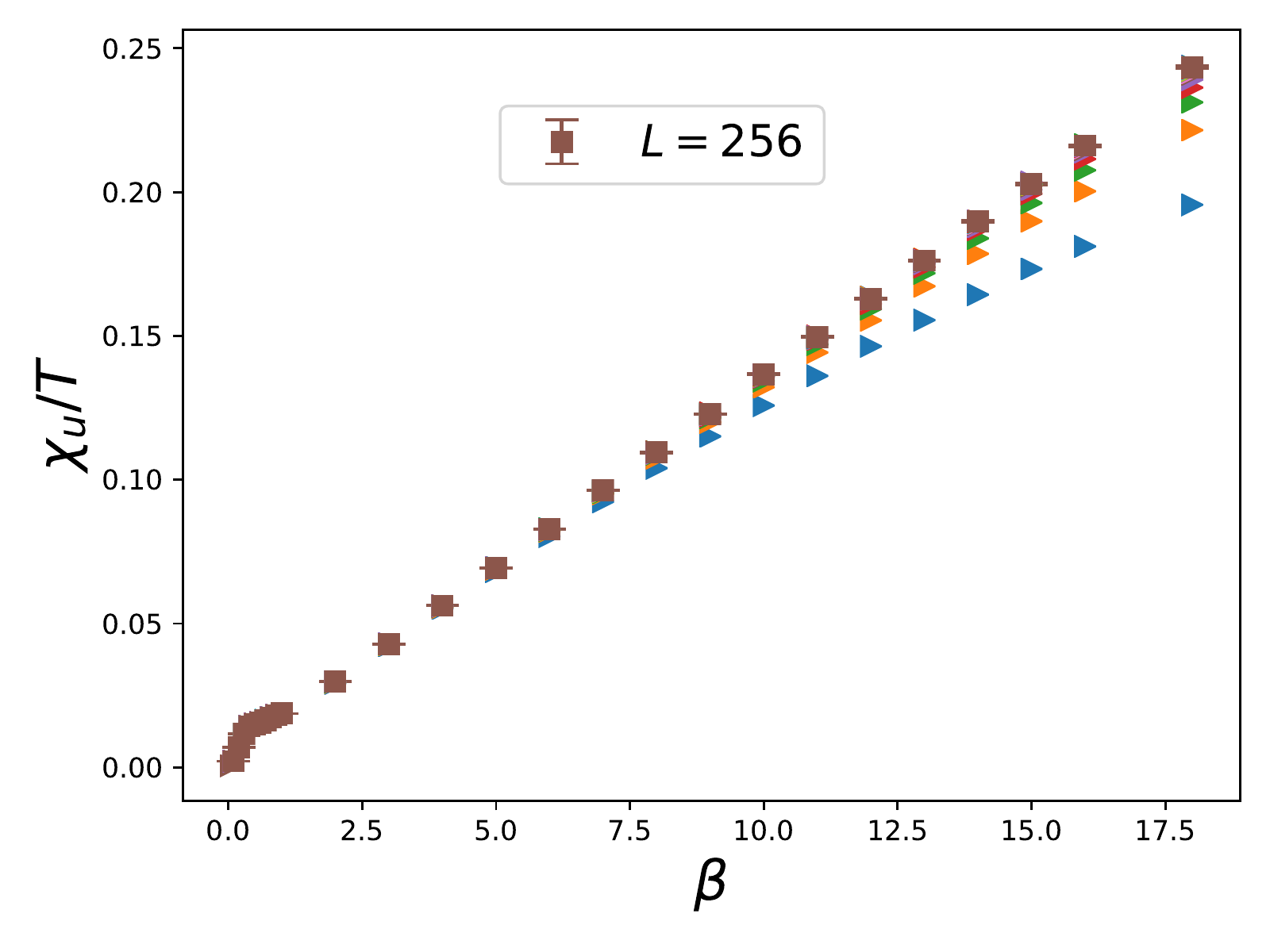}\vskip0.25cm
\end{center}\vskip-0.5cm
\caption{$\chi_u /T$ as a function
  of $\beta$ for various $L$ for the staggered $JQ_3$ models.}
\label{chiu2}
\end{figure}

In fig.~\ref{chiu2} $\chi_u/T$ is considered as a function of $\beta$ for the staggered $JQ_3$ model.
For a first order phase transition, one expects the magnitude of the winding number squared grows
with $\beta$.
Hence the rising of $\chi_u/T$ in the low temperature region is a signal of
first order phase transition for the staggered $JQ_3$ model. This is consistent with the conclusion established in the literature.

\begin{figure}
\vskip-0.5cm
\begin{center}
    \includegraphics[width=0.45\textwidth]{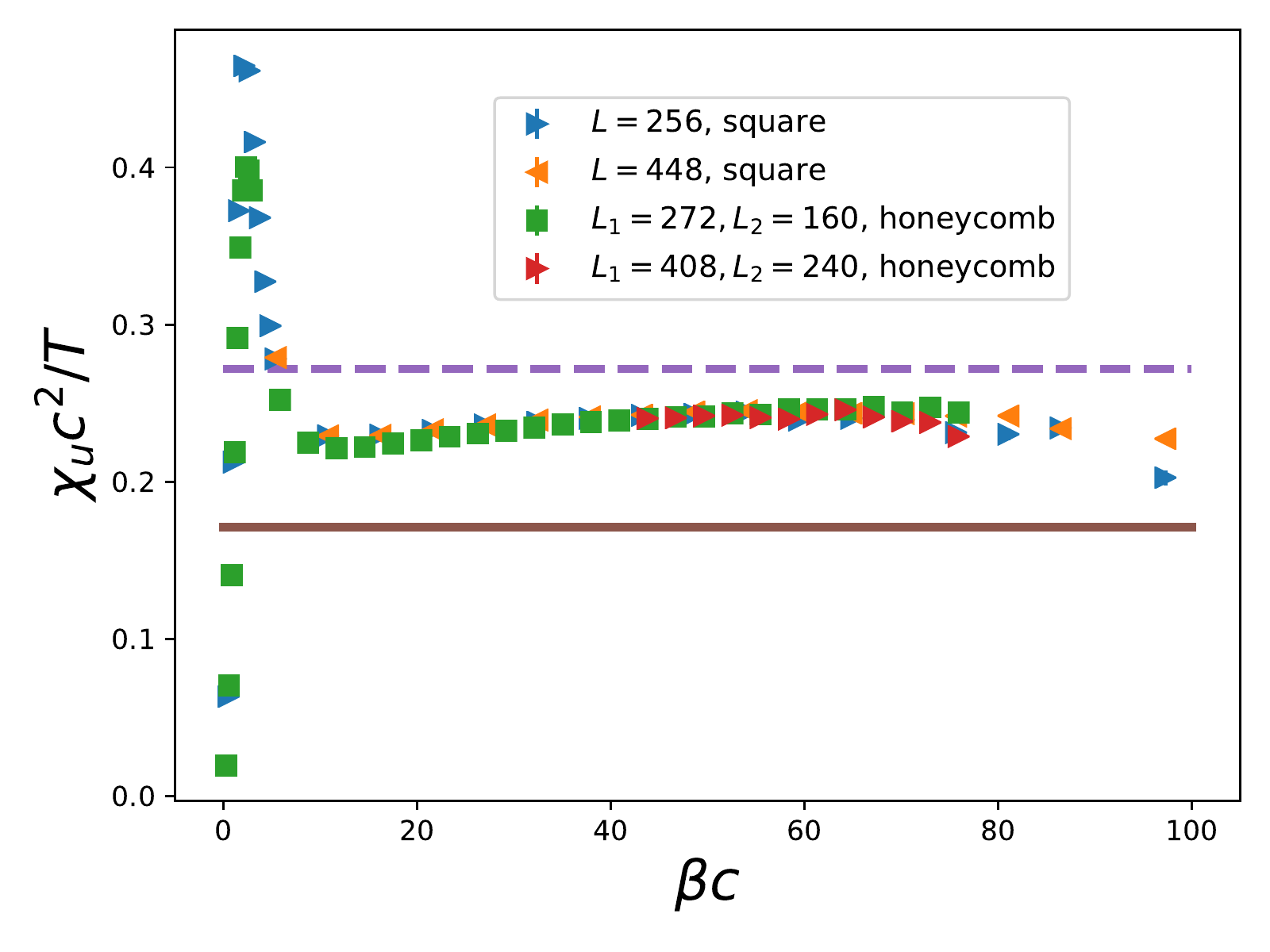}\vskip0.25cm
\end{center}\vskip-0.5cm
\caption{$\chi_u c^2/T$ as functions
  of $\beta c$ for various $L$ for the ladder and the honeycomb $JQ_3$ models.}
\label{chiu3}
\end{figure}

Similar to the outcomes associated with $S(\pi,\pi)$/$\left(\chi_s T\right)$,
the plateaus of $\chi_u c^2/T$ for both the ladder and the honeycomb $JQ_3$ models have the values of about the same magnitude,
and a good data collapse result is obtained if the data on large lattices of both models are considered as functions of $\beta c$,
again see fig.~\ref{chiu3}.
This observation can also be regarded as an evidence that both the transitions
of the ladder and the honeycomb $JQ_3$ model are likely continuous.

Interestingly, if the data of $\chi_u c^2/T$ of both the ladder and the honeycomb $JQ_3$ models are plotted as functions
of $\beta$, a slightly downgrade quality scaling than that of fig.~\ref{chiu3} appears as well.

Finally, it should be pointed out that the trend of moving downward for the quantity $\chi_u c^2/T$ at large values of $\beta$ shown
in figs.~\ref{chiu1} and \ref{chiu3} is due the observable $\chi_u$ and is an artifact. In particular, to obtain the correct bulk
$\chi_u$, taking the appropriate order of limits, namely, $\lim_{\beta \rightarrow \infty}\lim_{L \rightarrow \infty}$ is essential. 

\subsection{The spin stiffness $\rho_s$}

\begin{figure}
\vskip-0.5cm
\begin{center}
    \includegraphics[width=0.45\textwidth]{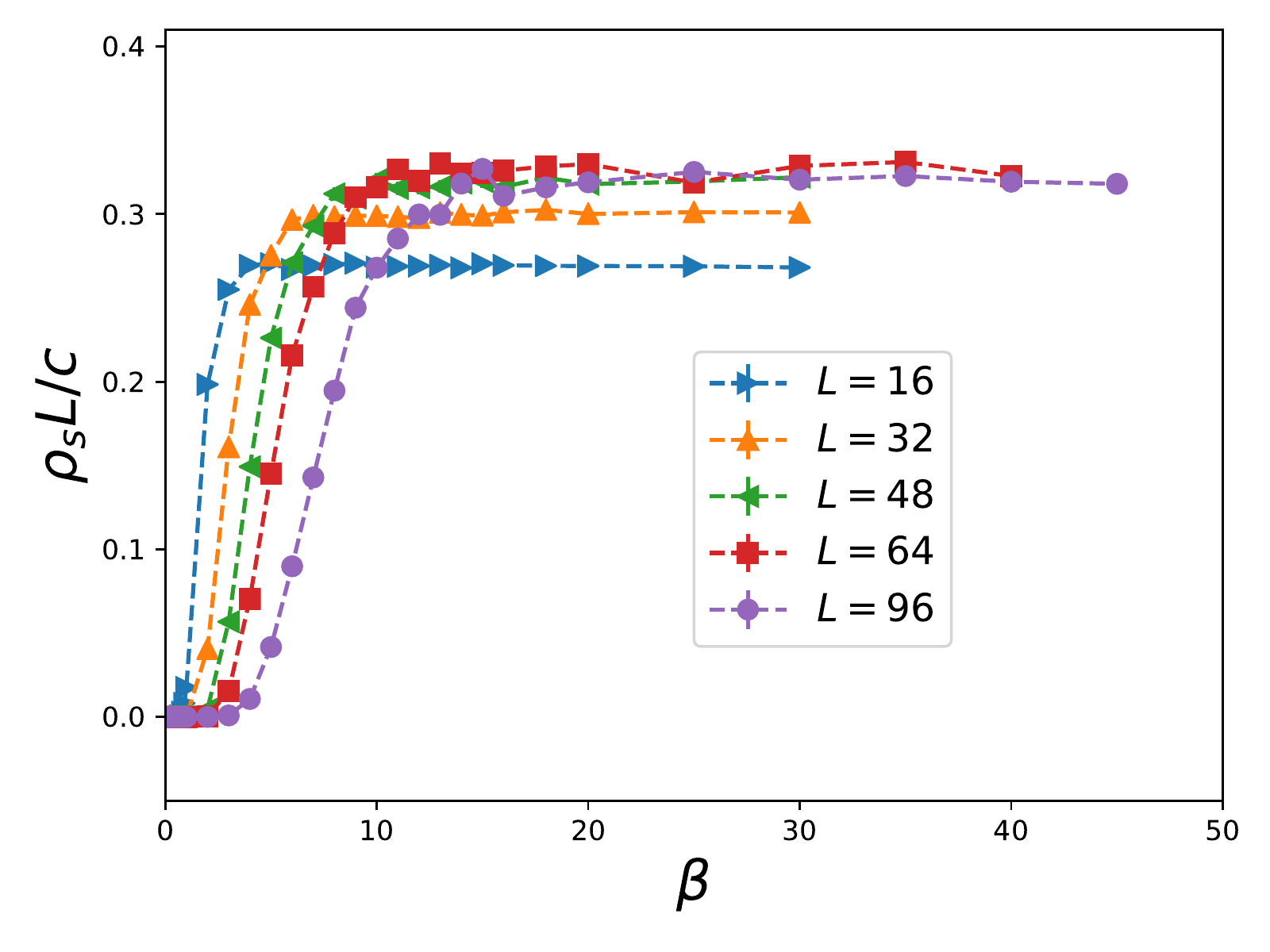}\vskip0.25cm
\end{center}\vskip-0.5cm
\caption{$\rho_sL /c$ as functions
  of $\beta$ for various $L$ for the ladder $JQ_3$ models.}
\label{rhosLc1}
\end{figure}

For the ladder $JQ_3$ model, the quantity $\rho_s L/c$ as functions of $\beta$ for several $L$ are
demonstrated in fig.~\ref{rhosLc1}. The results in fig.~\ref{rhosLc1} indicate that in the limit $LT \rightarrow \infty$, 
$\rho_s L/c$ approaches a value between 0.32 and 0.33. This number differ significantly from $0.18520$ calculated in
Ref.~\cite{Kau08}. However, if an alternative definition, namely $\rho_s = \frac{1}{2\beta}\left(\langle W_1^2\rangle + \langle W_2^2\rangle\right)$
is used, then our calculations lead to $\rho_sL/c \sim 0.215$ in the $LT \rightarrow \infty$ limit,
which is close to the predicted result of $0.18520$.

\begin{figure}
\vskip-0.5cm
\begin{center}
    \includegraphics[width=0.45\textwidth]{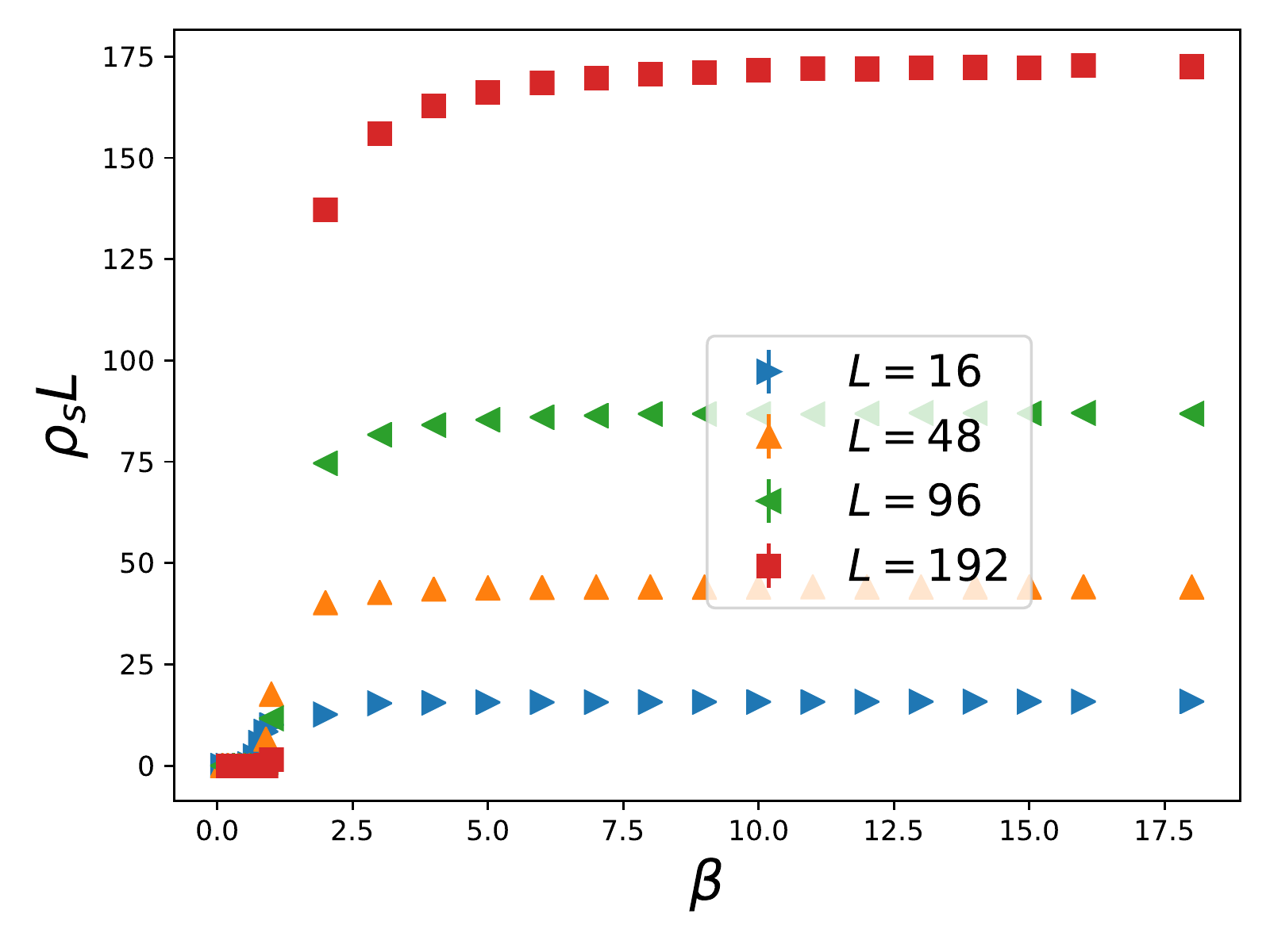}\vskip0.25cm
\end{center}\vskip-0.5cm
\caption{$\rho_sL /c$ as functions
  of $\beta$ for various $L$ for the staggered $JQ_3$ models.}
\label{rhosLc2}
\end{figure}

For the staggered $JQ_3$ model, it is anticipated that $\rho_s L/c$ will grow with $L$ (in the $LT \rightarrow \infty$ limit),
and this is indeed what's observed in our data, see fig.~\ref{rhosLc2}.

\begin{figure}
\vskip-0.5cm
\begin{center}
    \includegraphics[width=0.45\textwidth]{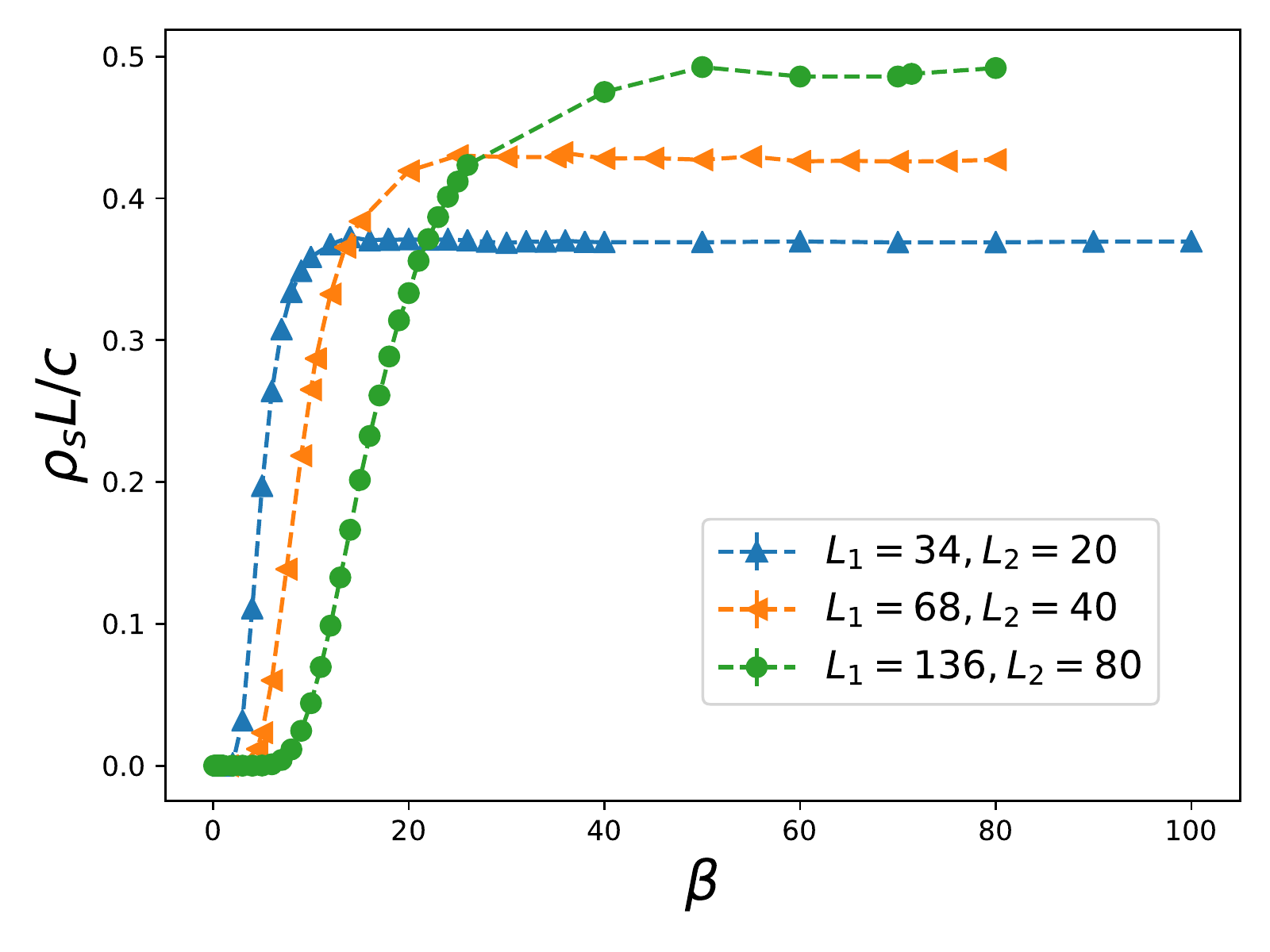}\vskip0.25cm
\end{center}\vskip-0.5cm
\caption{$\rho_sL /c$ as functions
  of $\beta $ for various $L$ for the honeycomb $JQ_3$ models.}
\label{rhosLc3}
\end{figure}

\begin{figure}
\vskip-0.5cm
\begin{center}
    \includegraphics[width=0.45\textwidth]{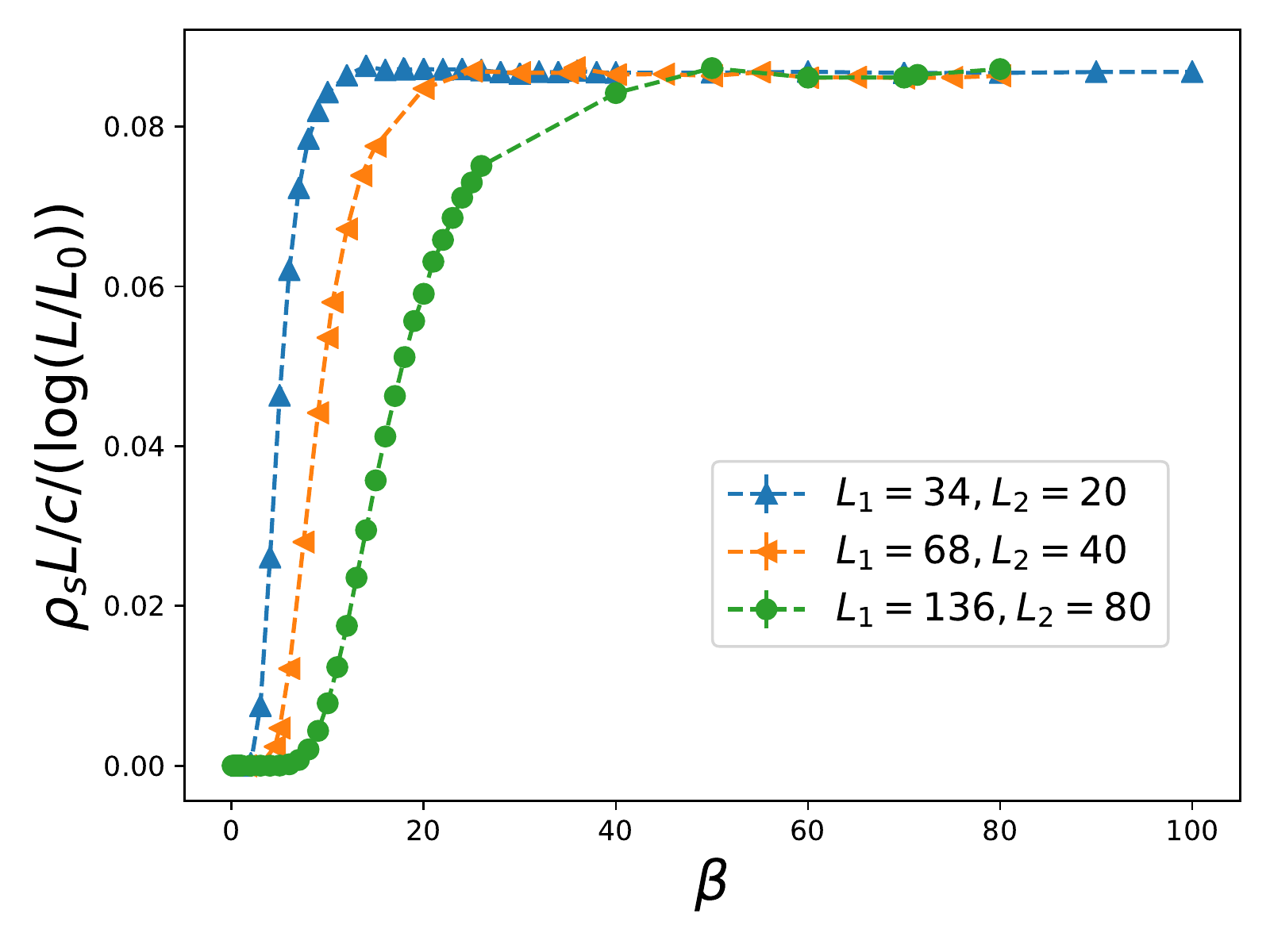}\vskip0.25cm
\end{center}\vskip-0.5cm
\caption{$\rho_sL /c/\left(\log\left(L/L_0\right)\right)$ as functions
  of $\beta $ for various $L$ for the honeycomb $JQ_3$ models. $L_0$ is 0.37.}
\label{rhosLc4}
\end{figure}

In fig.~\ref{rhosLc3}, $\rho_s L/c$ as functions of $\beta$ is shown for several honeycomb lattices.
Although no convergence (as $LT \rightarrow \infty$) is observed, it is interesting to notice
that a good data collapse emerges when a logarithmic correction is taken into account.
Specifically, if $\left(\rho_s L/\log\left(L/L_0\right)\right)/c, L_0 = 0.37$ is considered as functions of
$\beta c$, then a high quality scaling appears, see fig.~\ref{rhosLc4}. This scenario is consistent with
that obtained in Ref.~\cite{Puj13}. Finally, since
a free parameter is allowed in the logarithmic correction, namely one
can use $a \log\left(L/L_0\right)$ with $a$ being some constant instead of $\log\left(L/L_0\right)$,
we do not contrast the results related to the spin stiffness $\rho_s$
on the honeycomb lattice with that on the square lattice.

\section{Discussions and Conclusions}

Using the first principle quantum Monte Carlo simulations, we investigate the QCR of three 2D $SU(2)$ $JQ_3$ models on both
the square and the honeycomb lattices. Particularly, three universal quantities associated with QCR, namely $S(\pi,\pi)/\left(\chi_s T\right)$,
$\chi_u c^2/T$ and $\rho_s L /c$ are obtained and their $\beta$ ($\beta c$) dependence is studied.

Among the three studied models, the ladder and the honeycomb $JQ_3$ models are shown to possess second order phase
transitions from the N\'eel phase to the VBS phase \cite{Lou09,Puj13,Har13}, and for the staggered $JQ_3$ model the associated transition
is first order \cite{Sen10}. Our investigation related to the QCR leads to results
consistent with these scenarios.

Specifically, while for the ladder and the honeycomb $JQ_3$ models, 
the quantity $S\left(\pi,\pi\right)/\left(\chi_s T\right)$ ($\chi_u c^2/T$)
of both models reach the same plateau value and a good scaling emerges when considered as a function of
$\beta c$, no such behavior is found
for the staggered $JQ_3$ models. Moreover, $\rho_s L /c$ obtains a constant outcome in the $LT \rightarrow \infty$ limit for the
ladder model. Similar situation applies to the honeycomb model when a logarithmic correction
to the observable $\rho_s L /c$ is taken into account. For the staggered model both the quantities
$\chi_u c^2/T$ and $\rho_s L/c$ grow with $L$ or $\beta$. This is in agreement with the fact that the targeted phase transition
of the staggered model is first order.

It is surprising to observe that to obtain a good scaling, the universal quantity associated with $\rho_s$ on the honeycomb lattice
requires a logarithmic correction. This is not the case for the same observable on the square lattice. It will be interesting
to understand this from a theoretical perspective.

Apart from the three universal quantities considered here, one frequently studied
observable associated with QCR is the Wilson ratio $W = \chi_u T/C_V$. Here $C_V$ is the specific heat and can be
calculated directly or through the internal energy density $E$. However, due to the subtlety of both approaches for the
determination of $C_V$
shown in Refs.~\cite{Sen15,Pen20}, here we do not attempt to carry out a high precision calculation of the quantity $W$.
Nevertheless, the investigation conducted here provide evidence to support the scenario that for the ladder and the honeycomb
$JQ_3$ models, the associated quantum phase transitions from the N\'eel phase to the VBS phase are likely to be continuous.

The results presented here are not only interesting in themselves, but also can be used as criterions to distinguish second
order phase transitions from first order ones for exotic phase transitions such as the DQC studied here.

\vskip2.5cm
This study is partially supported by the Ministry of Science and Technology of Taiwan.

~
~
~

~
~
~
~
~
~
~
~
~
~
~
~
~


\begin{thebibliography}{99}


\bibitem{Nig92}
Nigel~Goldenfeld, {\it Lectures On Phase Transitions And The Renormalization
Group (Frontiers in Physics)} (Addison-Wesley, 1992).
  
\bibitem{Car96}
  J. Cardy, {\it Scaling and Renormalization in Statistical Physics}, Cambridge
  University Press, Cambridge, UK, 1996.

\bibitem{Car10}
Lincoln~D.~Carr, {\it Understanding Quantum Phase Transitions (Condensed Matter Physics)}
(CRC Press, 2010).

\bibitem{Sac11}
S.~Sachdev, {\it Quantum Phase Transitions } (Cambridge University Press, Cambridge, 2nd edition, 2011).

  \bibitem{San07}
    A. W. Sandvik, Phys. Rev. Lett. {\bf 98}, 227202 (2007).

 \bibitem{Lou09}
J. Lou, A. W. Sandvik, and N. Kawashima, Phys. Rev.
B. 80, 180414(R) (2009).

\bibitem{Sen10}
Arnab Sen and Anders W. Sandvik
Phys. Rev. B {\bf 82}, 174428 (2010).

\bibitem{Gin50}
V.L. Ginzburg and L.D. Landau, Zh. Eksp. Teor. Fiz. {\bf 20}, 1064 (1950). English translation in: L. D. Landau, Collected papers (Oxford: Pergamon Press, 1965) p. 546.


  \bibitem{Sen040}
T. Senthil, L. Balents, S. Sachdev, A. Vishmanath and
M. P. A. Fisher, Science {\bf 303} 1490 (2004).

\bibitem{Sen041}
T. Senthil, L. Balents, S. Sachdev, A. Vishwanath, and
M. P. A. Fisher, Phys. Rev. B {\bf 70}, 144407 (2004).


\bibitem{Kuk06}
Anatoly Kuklov, Nikolay Prokof'ev, Boris Svistunov, Matthias Troyer,
Annals of Physics
Volume 321, Issue 7, July 2006, Pages 1602-1621.

    \bibitem{Mel07}
    R. G. Melko and R. K. Kaul, Phys. Rev. Lett. {\bf 100},
017203 (2008).

\bibitem{Jia08}
 F.-J. Jiang, M. Nyfeler, S. Chandrasekharan, and U.-J.
Wiese, J. Stat. Mech. (2008) P02009.

\bibitem{Kuk08}

  A. B. Kuklov, M. Matsumoto, N. V. Prokof'ev, B. V. Svistunov, M. Troyer,
  Phys. Rev. Lett. {\bf 101}, 050405 (2008).

\bibitem{San10}
A. W. Sandvik, Phys. Rev. Lett. {\bf 104}, 177201 (2010).


\bibitem{Che13}
Kun Chen, Yuan Huang, Youjin Deng, A. B. Kuklov, N. V. Prokof’ev, and B. V. Svistunov
Phys. Rev. Lett. {\bf 110}, 185701 (2013).


\bibitem{Har13}
Kenji Harada, Takafumi Suzuki, Tsuyoshi Okubo, Haruhiko Matsuo, Jie Lou, Hiroshi Watanabe, Synge Todo, and Naoki Kawashima
Phys. Rev. B 88, 220408(R) (2013).

\bibitem{Sha16}
  Hui Shao, Wenan Guo, Anders W. Sandvik,
  Science 352, 213 (2016).

\bibitem{San201}
  Anders W. Sandvik, Bowen Zhao,
Chin. Phys. Lett. {\bf 37}, 057502 (2020)

\bibitem{Iin19}
Shumpei Iino, Satoshi Morita, Anders W. Sandvik, Naoki Kawashima,
 J. Phys. Soc. Jpn. 88, 034006 (2019).


\bibitem{Mer66}
N. D. Mermin and H. Wagner, Phys. Rev. Lett. {\bf 17}, 1133 (1966).

\bibitem{Hoh67}
P. C. Hohenberg, Phys. Rev. {\bf 158}, 383 (1967).

\bibitem{Col73}
Sidney Coleman, Communications in Mathematical Physics volume 31, pages 259–264 (1973).

\bibitem{Gel01}
  Axel Gelfert and Wolfgang Nolting, J. Phys.: Condens. Matter 13 R505 (2001).
 
\bibitem{Chu93}
A.~V.~Chubukov and S.~Sachdev, Phys. Rev. Lett. {\bf 71}, 169 (1993).

\bibitem{Chu931}
A. V. Chubukov and S. Sachdev, Phys. Rev. Lett. {\bf 71}, 2680 (1993).

\bibitem{Chu94}
A.~V.~Chubukov, S.~Sachdev, and J.~Ye, Phys. Rev. B {\bf 49}, 11919 (1994).

\bibitem{San95}
A.~W.~Sandvik, A.~V.~Chubukov, and S.~Sachdev, Phys. Rev. B {\bf 51}, 16483 (1995)

\bibitem{Tro96}
M.~Troyer, H.~Kantani, and K.~Ueda, Phys.~Rev.~Lett. {\bf 76}, 3822 (1996).

\bibitem{Tro97} 
Matthias Troyer, Masatoshi~Imada, and Kazuo~Ueda, J. Phys. Soc. Jpn. 66, 2957 (1997).

\bibitem{Tro98}
Jae-Kwon Kim and Matthias~Troyer, Phys.~Rev.~Lett. {\bf 80}, 2705 (1998).

\bibitem{Kim99}
Y.~J.~Kim, R.~J.~Birgeneau, M.~A.~Kastner, Y.~S.~Lee, Y.~Endoh, G.~Shirane, and K.~Yamada,
Phys. Rev. B 60, 3294 (1999).

\bibitem{Kim00}
Y.~J.~Kim and R.~J.~Birgeneau, Phys. Rev. B {\bf 62}, 6378 (2000).

\bibitem{San11}
A. W. Sandvik, V. N. Kotov, and O. P. Sushkov, Phys.
Rev. Lett. {\bf 106}, 207203 (2011).

\bibitem{Sen15}
A. Sen, H. Suwa, and A. W. Sandvik, Phys. Rev. B {\bf 92}, 195145 (2015).


\bibitem{Tan181}
	D.-R. Tan and F.-J. Jiang, Phys. Rev. B {\bf 98}, 245111 (2018).

  \bibitem{Kau08}
    Ribhu K. Kaul, Roger G. Melko, Phys. Rev. B {\bf 78}, 014417 (2008).

\bibitem{Kau081}    
Ribhu K. Kaul and Subir Sachdev, Phys. Rev. B {\bf 77}, 155105 (2008).
  \bibitem{Puj13}
Sumiran Pujari, Kedar Damle, and Fabien Alet,
Phys. Rev. Lett. {\bf 111}, 087203 (2013).

\bibitem{San99}
A.~W.~Sandvik, Phys. Rev. B {\bf 55}, R14157 (1999).

\bibitem{San101}
A.~W. Sandvik, AIP Conf. Proc. 1297, 135 (AIP, New York, 2010).



 \bibitem{Jia111}
F.-J. Jiang, Phys. Rev. B {\bf 83}, 024419 (2011).        

\bibitem{Jia112}
F.-J. Jiang and U.-J. Wiese,
Phys. Rev. B {\bf 83}, 155120 (2011).


\bibitem{Pen20}
Jhao-Hong Peng, L.-W. Huang, D.-R. Tan, and F.-J. Jiang,
Phys. Rev. B {\bf 101}, 174404 (2020).















\end{thebibliography}
\end{document}